\let\pdfoutput\defined 
\documentclass{PoS}
\usepackage{subfigure, epsfig, rotate, epsf,setspace, amssymb, amsxtra, placeins}
\newcommand{\msbar}{\overline{\mathrm{MS}}}
\newcommand{\phys}{\mathrm{phys}}

\newcommand{\res}{\mathrm{res}}

\title{Continuum results for light hadrons from 2+1 flavor DWF ensembles}

\ShortTitle{Continuum results for light hadrons from 2+1 flavor DWF ensembles}

\author{\speaker{Chris Kelly}, Peter Boyle%
	\\
        School of Physics\\
        The University of Edinburgh\\
	James Clerk Maxwell Building, King's Buildings\\
	Mayfield Road\\
	Edinburgh EH9 3JZ, UK \\
        E-mail: \email{christopher.kelly@ed.ac.uk}, \email{paboyle@ph.ed.ac.uk} }

\author{Chris Sachrajda
\\
School of Physics and Astronomy\\
University of Southampton\\
Southampton SO17 1BJ, UK \\
E-mail: \email{cts@soton.ac.uk} }

\author{For the RBC and UKQCD collaborations}

\abstract{From simultaneous fits to the data from the 2+1 flavor DWF ensembles generated by the RBC and UKQCD collaborations at two different lattice spacings, we present preliminary continuum results for light hadrons. We focus on light pseudoscalar decay constants and quark masses.  Several approaches to the calculation of the lattice spacing
are discussed and the errors associated with the chiral extrapolation are explained.  We make use of reweighting in the dynamical strange quark mass such that our ensembles have a self-consistently determined strange quark mass.}

\FullConference{The XXVII International Symposium on Lattice Field Theory \\
		 July 25 - 31, 2009\\
		 Beijing, China}

\begin{document}
\vspace{-1.5cm}
\section{Introduction}
\vspace{-0.3cm}
In these proceedings we discuss continuum physical results for light hadronic quantities obtained through a combined continuum extrapolation of the RBC and UKQCD collaboration's $32^3\times 64$ and $24^3\times 64$ domain wall fermion lattices with $L_s = 16$ and the Iwasaki gauge action at $\beta = 2.13$ and $2.25$ respectively. The lattice spacings, as determined by the combined analysis, are around $2.32$ GeV and $1.73$ GeV for the $32^3\times 64$ and $24^3\times 64$ lattices respectively, such that the lightest unitary pion masses are around $295$ MeV and $330$ MeV.

The layout of these proceedings is as follows: We first discuss scaling between the two lattices, matching at masses within the range of our simulations. We then discuss our simultaneous continuum and chiral extrapolation, determining the physical quark masses, lattice spacings and the continuum limit of the fitted observables. We make use of reweighting \cite{ChulwooProc2009} in the strange sea sector with a corresponding interpolation in the valence sector to reach a physical unitary strange quark mass. We then determine the neutral kaon mixing amplitude $B_K$ for which we quote a preliminary physical value.
\vspace{-5mm}
\section{Scaling at simulated mass points}
\vspace{-2mm}
We compare our two lattices by matching dimensionless quantities at a simulated (unphysical) mass point. The method is as follows. (i) we choose a match point on either lattice ($L_{\mathrm{match}}$), defined as a choice of simulated light quark mass $m_l$ and heavy quark mass $m_h$ (ii) we interpolate two dimensionless quantities $R_l$ and $R_h$ on the other lattice ($L_{\mathrm{interp}}$) until they match those on $L_{\mathrm{match}}$. These quantities are chosen to be ratios of lattice quantities which are sensitive to the desired quark mass. Examples include $R_l = M_{PS,ll}/M_\Omega$ and $R_h = M_{PS,hl}/M_\Omega$, where $M_{PS,ll}$ and $M_{PS,hl}$ are the unitary light-light and heavy-light pseudoscalar (PS) masses respectively. Prior to the interpolation, the numerator and denominator quantities of both $R_l$ and $R_h$ are fit linearly in $m_l$ and $m_h$, and the interpolation is performed by varying these parameters until the quantities match.

\begin{figure}[htb]
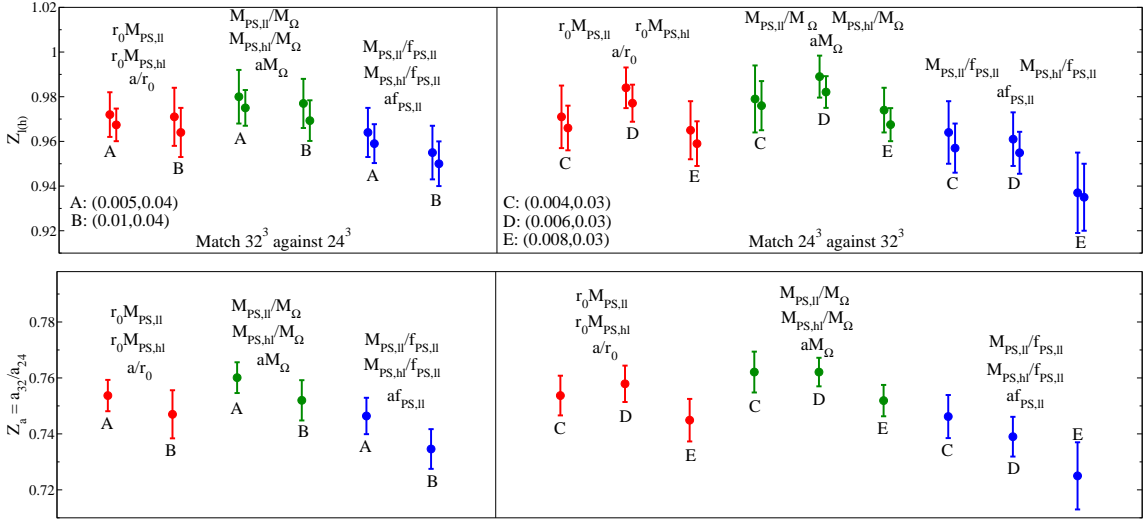

\vspace{-7mm}
\centering
\subfigure{
  \includegraphics[width=\textwidth]{zlfull.eps}
}\vspace{0.0cm}
\subfigure{
  \includegraphics[width=\textwidth]{zafull.eps}
}
\vspace{-3mm}
\caption{Plots of $Z_l$, $Z_h$ (top) and $Z_a$ (bottom) determined at several match points labelled A-E on both ensembles, using differing ratios $R_l$ and $R_h$ and $Q_a$ (labelled in order). In the right-hand panes, for which $L_{\mathrm{match}}$ and $L_{\mathrm{interp}}$ are swapped with respect to the left, we plot the inverse quantity $Z_{l/h/a}^{-1}$ for easier comparison.}
\label{zlzaplots}
\vspace{-4mm}
\end{figure}

Once the match point has been found, we can determine the ratio of the lattice spacings $Z_a$ by taking the ratio of a lattice quantity $Q_a$ (e.g. $aM_\Omega$) between the two lattices $Z_a = \frac{Q_a(L_{\mathrm{interp}})}{Q_a(L_{\mathrm{match}})}$. We can also determine the ratio of the physical quark masses
\vspace{-4mm}
\begin{equation}
Z_l = \frac{m_l^\phys(L_{\mathrm{interp}})}{m_l^\phys(L_{\mathrm{match}})} = \frac{1}{Z_a}\frac{\tilde{m}_l(L_{\mathrm{interp}})}{\tilde{m}_l(L_{\mathrm{match}})} \hspace{0.3cm},\hspace{0.3cm}Z_h = \frac{m_h^\phys(L_{\mathrm{interp}})}{m_h^\phys(L_{\mathrm{match}})} = \frac{1}{Z_a}\frac{\tilde{m}_h(L_{\mathrm{interp}})}{\tilde{m}_h(L_{\mathrm{match}})}\,,
\vspace{-3mm}
\end{equation}
where $\tilde{m}=m+m_\res$. Figure \ref{zlzaplots} shows $Z_l$, $Z_h$ and $Z_a$ for various choices of match point, $R_{l/h}$ and $Q_a$, showing good consistency between the choices.

\begin{figure}[htb]
\vspace{-0mm}
\centering
\includegraphics[width=\textwidth,bb=-60 33 1503 395]{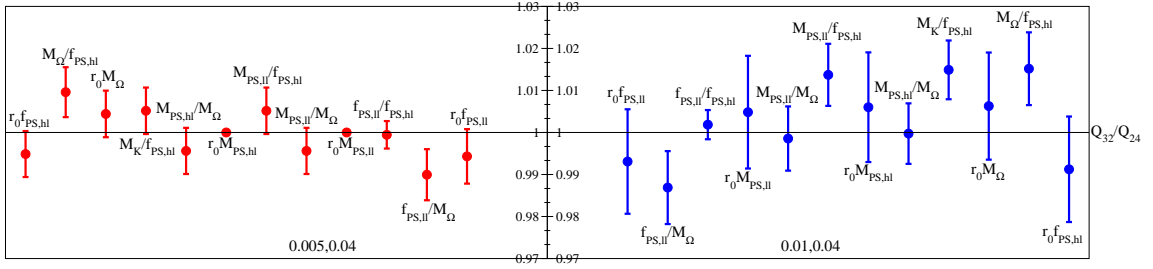} 
\vspace{-7mm}
\caption{A comparison of different dimensionless quantities $Q$ between the lattices at two match points $(m_l,m_h)=(0.005,0.04)$ (left) and $(0.01,0.04)$ (right) on the $24^3\times 64$ lattice. $Z_l$ and $Z_h$ are calculated using $R_l = r_0 M_{PS,ll}$ and $R_h = r_0 M_{PS,hl}$. $Q_{32}/Q_{24}$ being unity defines the line of perfect scaling.}
\label{scalingplot}
\vspace{-6mm}
\end{figure}

Picking a representative $Z_l$ and $Z_h$ we can determine the matching quark masses for any given simulated point on  $L_{\mathrm{match}}$. We compare dimensionless ratios of other quantities over a number of match points in order to demonstrate the scaling of our data. Figure \ref{scalingplot} shows two match points on the $24^3\times 64$ lattice, demonstrating scaling violations of less than $2\%$. Here the Wilson parameter $r_0$ was obtained as per ref. \cite{Antonio:2006px}.

\vspace{-3mm}
\section{Simultaneous continuum fits}
\vspace{-2mm}
In order to make the best use of our data, we perform simultaneous fits to both lattices. We fit all quantities in physical units in a fixed renormalisation scheme. The renormalisation constants for the light and heavy quarks can be left as fit parameters or can be frozen to a predetermined value. For this analysis we assume the light and heavy quark renormalisation constants are equal, which is justified by the results shown in figure \ref{zlzaplots}. This is also expected from renormalised field theory, which predicts that a single fermion mass rescaling is needed to relate theories with different cutoff scales. We fix $Z_{32} = 1$ and $Z_{24} =\frac{m_l^\phys(32^3)}{m_l^\phys(24^3)} = Z_l$ from the scaling analysis of the previous section. This is equivalent to a non-canonical choice of renormalisation scale which can of course be corrected to $\msbar$ at a later stage. The lattice spacings (and physical quark masses) are self-consistently determined by fixing a set of quantities (detailed below) in the continuum limit to their known physical values within an iterative procedure.

We simultaneously perform a partially-quenched (PQ) fit to the following quantities: $f_{PS,xy}$, $f_{PS,hy}$, $M_{PS,xy}$, $M_{PS,hy}$ and $M_\Omega$. Here $m_x$ and $m_y$ are PQ light valence quark masses. Following our standard procedure \cite{Allton:2008pn} we include only light PS data for which the corresponding pion mass is less than $\sim420$ MeV. In order to allow for a continuum extrapolation, the fit forms are obtained through a double expansion in ChPT and $a^2$, taking terms of order $m_q a^2$ as higher order such that only the LO LEC terms have a lattice spacing dependence. The PS quantities are fit to NLO $SU(2)$ PQChPT fit forms and $M_\Omega$ is fit to a form linear in the light quark mass. For example, the NLO $SU(2)$ PQChPT fit form for the light PS decay constant in the case of degenerate valence quarks is
\vspace{-4mm}
\begin{equation}f_{PS,xx} = f(1+c_f^a a^2) - \frac{2(\chi_x+\chi_l)}{(32\pi^2 f)}\log\left(\frac{\chi_x+\chi_l}{2\Lambda_\chi^2}\right) + \frac{16}{f}L_4 \chi_l + \frac{4}{f}L_5\chi_x\,,\vspace{-3mm}\end{equation}
where $\chi_q = 2BZ_lm_q$. The fits are performed at the physical $m_h$ as determined self-consistently through an iterative procedure with the physical kaon mass as input, using reweighting to iterate the sea strange mass while interpolating in the valence sector. The physical $m_l$ is obtained by inverting the pion fit form on the physical pion mass. The two lattice spacings are similarly fixed by requiring that the predicted continuum $M_\Omega$ matches its experimental value and that the ratio of lattice spacings is equal to that determined from the scaling analysis of the previous section.

\begin{figure}[tbp]
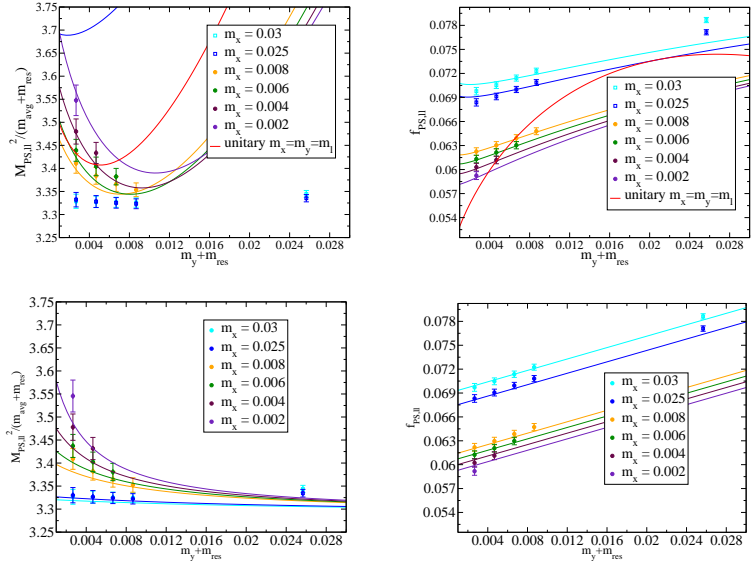

\vspace{-7mm}
\centering
\subfigure{ 
\includegraphics[width=0.30\textwidth]{plot_mps_over_mavg_latt0_ml0.004_pretty.eps}
\label{globalchptfitsa}  
}
\hspace{0.02\textwidth}
\subfigure{
\includegraphics[width=0.30\textwidth]{plot_fps_latt0_ml0.004_pretty.eps}
\label{globalchptfitsb}  
}\\
\vspace{0.00cm}
\subfigure{ 
\includegraphics[width=0.30\textwidth]{plot_mps_over_mavg_latt0_ml0.004_pretty_linear.eps}
\label{globalanalyticfitsa}
}
\hspace{0.02\textwidth}
\subfigure{
\includegraphics[width=0.30\textwidth]{plot_fps_latt0_ml0.004_pretty_linear.eps}
\label{globalanalyticfitsb}
}
\vspace{-3mm}
\caption{Results for $M_{PS,xy}$ (left) and $f_{PS,xy}$ (right) on the $32^3\times 64$ $m_l=0.004$ ensemble obtained from global fits using NLO PQChPT (upper) and leading-order, linear analytic expansion (lower) fit forms. Unfilled symbols indicate data that is not included in the fit. The curves show the fit form at the parameters equal to the data set of the same colour. The red curve is the unitary mass curve at which $m_x=m_y=m_l$. Note that the curvature apparent in the lower-left pane is a consequence of plotting  $2M_{PS,xy}^2/(\tilde{m}_x + \tilde{m}_y)$ against only a single term of the denominator; $\tilde{m}_y$.}
\vspace{-4mm}
\end{figure}

Figs. \ref{globalchptfitsa} and \ref{globalchptfitsb} show the NLO PQChPT fits for $M_{PS,xy}$ and $f_{PS,xy}$, not including finite volume (FV) corrections. Extrapolating to the continuum limit, we predict $f_\pi = 0.119(3)$ GeV and $f_K = 0.147(3)$ GeV. Comparing to the known physical values we see that these are $\sim 8\%$ ($\sim3\sigma$) and $\sim 5\%$ ($\sim2\sigma$) too low respectively. The inclusion of FV effects on $f_\pi$ obtained using FVPQChPT \cite{Allton:2008pn}, provides an upwards shift by $\sim 2.5\%$, which is insufficient to account for the difference. However, this disparity is of the order expected for the NNLO ChPT contributions obtained by squaring the NLO contribution, suggesting that a full NNLO fit is required to correctly reproduce the physical point. An investigation of the full NNLO fit is detailed by Mawhinney \cite{BobProc2009}, with the conclusion that these fits are not stable when applied to our data.


Although our final fits are uncorrelated, we also performed correlated fits using covariance matrices estimated by including increasing numbers eigenvectors \cite{CDawsonProc2009} finding no significant change in our results within our limited ability to estimate the covariance matrix.

\begin{figure}[tbp]
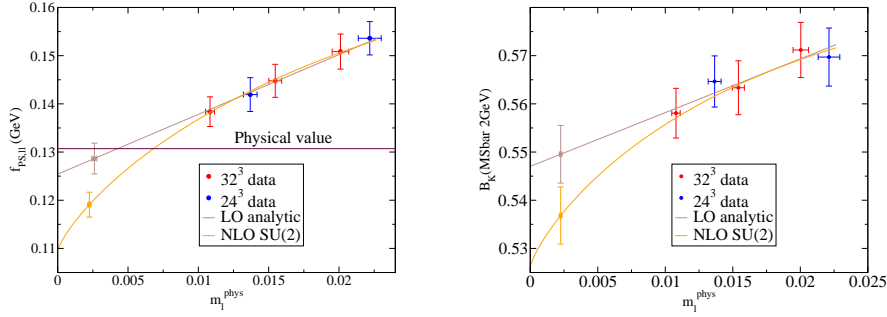

\vspace{-7mm}
\centering
\subfigure{
\includegraphics[width=0.34\textwidth]{fpi_unitary_flavour_su2_comparison.eps}
\label{fpicompare}
}\hspace{0.05\textwidth}
\subfigure{
\includegraphics[width=0.34\textwidth,bb=-18 -22 767 605]{bk_unitary_flavour_su2_comparison.eps}
\label{bkcontinuumcomparison}
}
\vspace{-3mm}
\caption{Plots of our $f_\pi$ (left) and $B_K$ (right) data on both lattices corrected to the continuum limit and the physical strange quark mass, overlayed with the NLO ChPT fit results (orange) and the LO analytic fit results (grey). On the former the physical value of $f_\pi$ is marked by a horizontal bar.}
\vspace{-2mm}
\end{figure}

We also detail an investigation using a form of analytic expansion about a non-zero unphysical pion mass, as advocated by Lellouch \cite{Lellouch:2009fg}. Using this approach we lose the ability to take the chiral limit and only extrapolate to the non-zero physical point. In figs. \ref{globalanalyticfitsa} and \ref{globalanalyticfitsb} we show the results for $M_{PS,xy}$ and $f_{PS,xy}$ obtained using only the leading order (linear) terms of the analytic expansion. As an example, the LO analytic form for $f_{PS,xy}$ is 
\vspace{-3mm}
\begin{equation}f_{PS,xy} = f_c(1+f_a a^2)+f_l Z_l m_l + f_v Z_l (m_x+m_y)/2\,.\vspace{-3mm}\end{equation}
These fit forms provide a better description of the data in our mass range than the NLO ChPT forms (c.f. figs. \ref{globalchptfitsa} and \ref{globalchptfitsb}) even to masses significantly higher than the $420$ MeV chiral cutoff. Figure \ref{fpicompare} shows a comparison between the two fit forms (without FV corrections) on the approach to the continuum limit. The LO analytic fit gives predicts $f_\pi = 0.129(3)$ GeV and $f_K = 0.151(3)$ GeV which agree very well with the known physical values.


We note that fitting an analytic expansion to higher masses and continuing this down into a region with known chiral non-analytic terms is not without risk, even though it worked surprisingly well for $f_\pi$. With this in mind we continue to take the central value from NLO ChPT and use these analytic fits to estimate a robust error: Based on the observation that the NLO corrections to $f_\pi$ are negative in sign and assuming the convergence of ChPT we conclude that the LO analytic fits are likely a good method for estimating the upper bound on the systematic errors for quantities such as $B_K$ for which there is no experimental value. Results at lighter masses are surely required to bring down the errors estimated by the spread of these two methods.

%

\begin{figure}[tp]
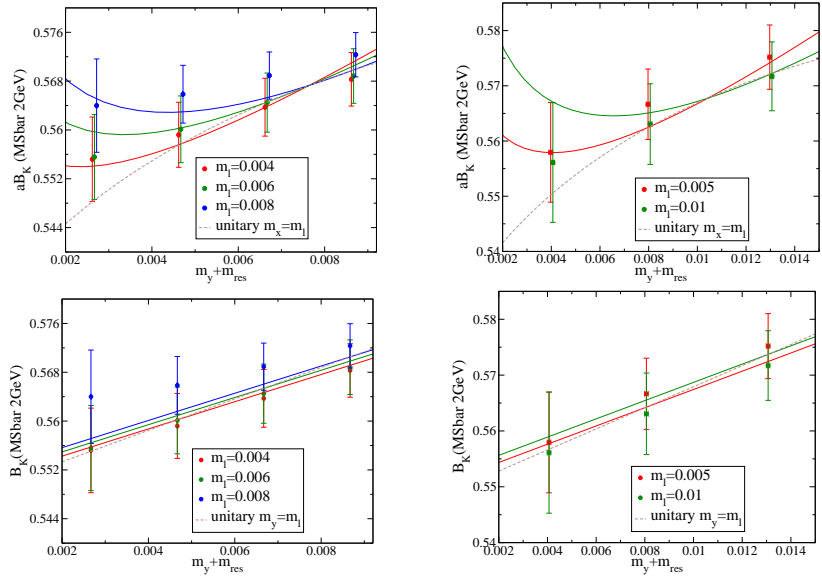

\vspace{-0mm}
\centering
\subfigure{
\includegraphics[width=0.32\textwidth]{plot_bk_32c.eps}
}\hspace{0.05\textwidth}
\subfigure{
\includegraphics[width=0.32\textwidth]{plot_bk_24c.eps}
}\\
\vspace{-2mm}
\subfigure{
\includegraphics[width=0.32\textwidth]{plot_bk_32c_flavour.eps}
}\hspace{0.05\textwidth}
\subfigure{
\includegraphics[width=0.32\textwidth]{plot_bk_24c_flavour.eps}
}
\vspace{-3mm}
\caption{A comparison of simultaneous $SU(2)$ PQChPT fits (upper) and LO analytic fits (lower) to the $B_K$ data on the $32^3$ (left) and $24^3$ (right) lattices reweighted to the physical strange quark mass.}
\label{bkfitcomparison}
\vspace{-2mm}
\end{figure}
%
%
\vspace{-4mm}
\section{The continuum limit of $B_K$}
\vspace{-2mm}
The neutral kaon mixing amplitude $B_K$ is defined as $B_K \equiv \frac{\langle K^0 | \mathcal{O}_{VV+AA} | \bar{K}^0 \rangle}{\frac{8}{3}f^2_K M^2_K}$, where the four-quark operator $\mathcal{O}_{VV+AA}$ is defined as
$\mathcal{O}_{VV+AA} \equiv (\bar s\,\gamma_\mu d)(\bar s\,\gamma_\mu d) + (\bar s\,\gamma_5\gamma_\mu d)(\bar s\,\gamma_5\gamma_\mu d)$. Using the operator product expansion, $B_K$ can be perturbatively related to $\epsilon_K$, the measure of indirect CP-violation in the kaon sector\cite{Buras:1998raa}. The relationship contains the CKM matrix phase $\delta$ which parameterises CP-violation in the Standard Model. $\epsilon_K$ is known precisely from measurements of $K_L\rightarrow\pi\pi$ and $K_S\rightarrow\pi\pi$ decays, thus the measurement of $B_K$ places constraints on the CP-violating phase. 

We fit $B_K$ on both lattices simultaneously to the $SU(2)$ PQChPT fit form
\vspace{-3mm}
\begin{equation}
B_K = B^0_K \Big{[}\,1 + c_a a^2 + \frac{c_0\chi_l}{f^2} + \frac{\chi_y c_1}{f^2} - \frac{\chi_y}{32\pi^2 f^2}\log\left(\frac{\chi_y}{\Lambda_\chi ^2}\right)\,\Big{]}\,,
\vspace{-3mm}
\end{equation}
where the kaon has been coupled into the theory at leading order in the relativistic expansion. Here $c_0$ and $c_1$ are LECs and again we have included a term $c_a$ parameterising the $\mathcal{O}(a^2)$ corrections. We simplify the fit by freezing the LECs $B$ (in $\chi_q$) and $f$ to the values obtained from the global fit analysis. As discussed in the previous section, we also perform a leading order analytic expansion fit to the data in order to estimate the systematic error on the fit function. The fits are performed to renormalised lattice quantities. We determine $Z_{B_K}$ using multiple non-exceptional momentum renormalisation conditions, which suppress chirality mixing. We used volume sources for good statistical precision. The conversion to $\msbar$ has been calculated by RBC-UKQCD to one loop for each of the schemes, and comparison of these gives a robust estimate of higher order errors (c.f. refs. \cite{Sturm:2009kb} and \cite{bkpaper2009}).

Figure \ref{bkfitcomparison} shows a comparison between the ChPT and analytic fits to our data at the physical strange quark mass. Here, the data on both lattices appears to display a slight non-monotonicity in the light quark mass, which on the $24^3$ lattices reverses the mass-ordering of the data with respect to that on the $32^3$. This effect is sub-statistical and we expect it to be resolved by an increase in the number of measurements. Figure  \ref{bkcontinuumcomparison} shows a comparison between the unitary ($m_y=m_l$) extrapolations of the data in the continuum limit using ChPT and LO analytic fit forms. This figure encapsulates our philosophy that the extrapolation may turn down or it may not, we simply do not know but we ensure that our errors cover this difference. Here the difference between the two physical points is much smaller than that for $f_\pi$, around $2.4\%$, but is again of a similar magnitude to the expected NNLO ChPT contributions.
\begin{table}[tbp]
\vspace{-1mm}
\centering
\footnotesize{
\begin{tabular}{c|cc|cc}
\hline\hline
Source & $24^3$ Mag. & \% err & New Mag. & \% err.\\
\hline	
stat 		 & 0.010 & 1.9  & 0.006  & 1.1\\
ChPT		 & 0.010 & 1.9 & 0.013   & 2.4\\
FV 		 & 0.005 & 1.0 &?? 	 & ?? \\
NPR 		 & 0.013 & 2.5 &0.013    & 2.4\\
Scaling 	 & 0.021 & 4.0 &0 	 & 0.0\\
Unphys. $m_h$ 	 & 0.005 & 1.0 &0 	 	 & 0.0\\
\hline
Total 		 & 0.030 & 5.7 &0.019         & 3.5\\
\hline
\end{tabular}
}
\caption{A breakdown of the quoted error on $B_K$ (right) compared to that of our previous determination\protect\cite{Antonio:2007pb} (left).}
\label{bkerrbreakdown}
\vspace{-5mm}
\end{table}

Using the lattice spacings and quark masses obtained from our global fit procedure we quote a preliminary physical value of $B_K^{\msbar}(2\,\mathrm{GeV}) = 0.537(19)$, not including finite-volume effects. These results will be published in our forthcoming paper \cite{bkpaper2009}. Comparing this to our previous value of $B_K^{\msbar}(2\,\mathrm{GeV}) = 0.524(30)$\cite{Antonio:2007pb} we see a consistent result with a substantially improved total error. Table \ref{bkerrbreakdown} contains a breakdown of the total error comparing to our previous result. Note that with the inclusion of reweighting in the heavy quark mass and by including $\mathcal{O}(a^2)$ corrections in our fit forms, systematic errors associated with the unphysical $m_h$ and with scaling violations have been removed and replaced with contributions to the statistical error. 
\vspace{-5mm}
\section{Conclusions and Acknowledgements}
\vspace{-2mm}
By matching the data on our $24^3\times 64$ and $32^3\times 64$ lattices at an unphysical match point, we have shown that our data is scaling to within $2\%$ for all quantities we tested. With confidence in our data we then described a global fit procedure whereby the quantities $M_{PS,ll}$, $f_{PS,ll}$, $M_{PS,hl}$, $f_{PS,hl}$ and $M_\Omega$ are all fit simultaneously on both lattices to NLO ChPT fit forms including a parameterisation of the leading $\mathcal(a^2)$ effects. This global fit is iterated within a procedure that determines the physical quark masses and lattice spacings. We make use of reweighting in the strange sea quark mass in order to reach the physical point. We show that NLO ChPT fit forms do not reproduce the known continuum value of $f_\pi$, giving a value around $5\%$ ($\sim 2\sigma_{\mathrm{stat}}$) too low after finite volume effects are included. We demonstrate that an analytic expansion (as advocated by Lellouch \cite{Lellouch:2009fg}) reproduces the physical $f_\pi$ and provides a better fit to the data at mass scales above the chosen chiral cutoff. Noting the risk of extrapolating these analytic forms down into a region with known chiral non-analyticities, we choose to continue to take the NLO ChPT central value for the subsequent analysis of the neutral kaon mixing parameter $B_K$, taking the LO analytic fits to this quantity as an upper bound on the systematic errors. We emphasise that results at lighter masses are necessary to reduce these errors. Finally we give a preliminary continuum result for $B_K$ as $B_K^{\msbar}(2\,\mathrm{GeV}) = 0.537(19)$, where the $3.5\%$ quoted error includes all systematics bar the finite-volume corrections which have not yet been calculated. We use the renormalisation parameter $Z_{B_K}$ calculated using multiple non-exceptional momentum schemes with volume sources for good statistical precision. The $\msbar$ conversion factors have all been calculated to one loop giving a robust systematic error estimate. We compare our $B_K$ value to our previous published result \cite{Antonio:2007pb}.

We thank all members of the UKQCD and RBC collaboration. Computations were performed on the QCDOC machines at the University of Edinburgh and Columbia University, the US DOE and RBRC facilities at the Brookhaven National Laboratory and at the Argonne Leadership Class Facility. The resources made available at the Argonne Leadership Class Facility were essential to the generation of our $32^3\times 64$ ensembles. The author was supported by the UK STFC.

\end{document}